\documentclass[usenatbib]{mnras}
\usepackage{amsmath}
\usepackage{amssymb}
\usepackage{graphicx}
\usepackage[utf8]{inputenc}
\usepackage{listings}
\usepackage{cleveref}
\usepackage{geometry}
\usepackage{ulem}
\usepackage{xcolor}
\usepackage{epsf}
\usepackage{epstopdf}
\usepackage{tikz}
\usetikzlibrary{arrows}

\DeclareMathOperator{\sech}{sech}

\newcommand{\change}[1]{{#1}}

\newcommand{\lkhd}{{\cal L}}

\newcommand{\redmapper}{redMaPPer}

\crefname{section}{\S\hspace{-0.3em}}{\S\S}
\Crefname{subsection}{\S\hspace{-0.3em}}{\S\S}

\title[Modeling Projected Velocity Distribution of Galaxies in DM Halos]{Accurate Model of the Projected Velocity Distribution of Galaxies in Dark Matter Halos}

\author[Aung et al.]{Han Aung$^{1}$, Daisuke Nagai$^{1}$, Eduardo Rozo$^{2}$, Brandon Wolfe$^{2}$, Susmita Adhikari$^{3,4}$ 
\\
$^{1}$Department of Physics, Yale University, 
New Haven, CT 06520, USA \\
$^{2}$Department of Physics, University of Arizona, Tucson, AZ 85721 USA\\
$^{3}$Indian Institute of Science Education and Research, Pune, Maharashtra, 411008, India\\
$^{4}$Department of Astronomy and Astrophysics, University of Chicago, Chicago, IL 60637, USA\\
}

\def\r200m{r_{\rm 200m}}
\def\rout{r_{\rm edge}}
\def\rta{r_{\rm ta}}

\def\sorb{\sigma_{\rm orb}}
\def\sinf{\sigma_{\rm inf}}
\def\sbg{\sigma_{\rm bg}}

\def\rratio{k_{\rm t}}

\def\Rout{R_{\rm edge}}
\def\Rorb{R_{\rm orb}}
\def\Rinf{R_{\rm inf}}
\def\Routp{R_{\rm p,edge}}
\def\Rorbp{R_{\rm p,orb}}
\def\Rinfp{R_{\rm p,inf}}

\def\surfaceorb{\Sigma_{\rm orb}}
\def\surfaceinf{\Sigma_{\rm inf}}
\def\surfacebg{\Sigma_{\rm bg}}
\def\surfaceda{\Sigma_{\rm da}}
\def\surfaceall{\Sigma_{\rm all}}

\hypersetup{draft}
\begin{document}

\Crefname{equation}{Eq.}{Eqs.}
\Crefname{figure}{Fig.}{Figs.}

\maketitle
\begin{abstract}
We present a percent-level accurate model of the line-of-sight velocity distribution of galaxies around dark matter halos as a function of projected radius and halo mass.  The model is developed and tested using synthetic galaxy catalogs generated with the UniverseMachine run on the Multi-Dark Planck 2 N-body simulations.  The model decomposes the galaxies around a cluster into three kinematically distinct classes: orbiting, infalling, and interloping galaxies.  We demonstrate that: 1) we can statistically distinguish between these three types of galaxies using only projected line-of-sight velocity information; 2) the halo edge radius inferred from the line-of-sight velocity dispersion is an excellent proxy for the three-dimensional halo edge radius; 3) we can accurately recover the full velocity dispersion profile for each of the three populations of galaxies. Importantly, the velocity dispersion profiles of the orbiting and infalling galaxies contain five independent parameters --- three distinct radial scales and two velocity dispersion amplitudes --- each of which is correlated with mass.  Thus, the velocity dispersion profile of galaxy clusters has inherent redundancies that allow us to perform nontrivial systematics checks from a single data set. We discuss several potential applications of our new model for detecting the edge radius and constraining cosmology and astrophysics using upcoming spectroscopic surveys.
\end{abstract}

\section{Introduction}

The abundance of galaxy clusters as a function of mass makes them powerful cosmological probes.  One way to infer cluster masses is using galaxy dynamics \citep{Evrard08,Bocquet15}. However, the utility of cluster samples can be limited by systematic uncertainties associated with nonlinear cluster astrophysics \citep{allen11,Pratt19}. One of the new frontiers in cluster cosmology
exploits the outskirts of galaxy clusters, where the impacts of poorly understood baryonic effects are modest \citep{Walker19}. Recent studies show that the phase space information around and outside the galaxy clusters enables dynamical mass estimation from larger regions around the cluster \citep{Hamabata19}, can place constraints on modified gravity \citep{Lam12, Zu2014}, and enable the use of halo boundaries as a standard ruler \citep{Wagoner2020}. 

Recent work demonstrated that the phase space structure of the halo exhibits two different populations of galaxies: 1) orbiting galaxies that have been inside the cluster; and 2) infalling galaxies which have never been inside the cluster \citep{Aung2020}. Moreover, there is a bonafide edge radius beyond which no orbiting galaxies can be found \citep{Bakels2021}. This edge radius provides a better definition of halo radius than traditional overdensity definitions: it denotes the halo boundary within which all orbiting dark matter particles and subhalos reside. The halo edge radius is intimately related to the traditional splashback radius.  While the splashback radius was originally defined in terms of the steepest slope of the halo's density profile \citep{diemer_kravtsov2014,adhikari_etal2014,more_etal2015}, modern definitions rely on a specific percentile of the distribution of orbiting particle apocenters, typically in the range $75\%$ to $87\%$ so as to match the ``steepest slope'' definition \citep{diemer_etal17}. The halo edge radius is the smallest possible 100-percentile splashback radius.

The edge radius of galaxy clusters has been detected using the Sloan Digital Sky Survey \redmapper\ cluster catalog \citep{Tomooka2020}. Specifically, the edge radius is seen as a ``break'' in the velocity dispersion profile of galaxy clusters.  Because the amplitude of the velocity dispersion profile must be correlated with the halo edge radius (more massive halos are bigger), one can use the halo edge radius as a standard ruler \citep{Wagoner2020}.  A measurement of the amplitude of the velocity dispersion profile allows us to infer the halo edge radius in Mpc, while measurements of the profile as a function of angle allow us to determine the angle spanned by the halo edge radius.  Together, these two pieces of data allow us to measure the distance to galaxy clusters.
Forecasts show that the next generation of spectroscopic surveys such as Dark Energy Spectroscopic Instrument \citep[DESI,][]{desi} will provide enough statistics to measure the Hubble constant with percent level precision using this technique \citep{Wagoner2020}. 

Actually achieving a percent level measurement of the Hubble constant requires controlling the theoretical and observational systematics impacting the measurement with sub-percent level precision \citep{Wagoner2020}.  Current major sources of uncertainties in measuring line-of-sight velocity dispersion profile are: 1) accurate modeling of the nonlinear dynamics of dark matter particles in the infall and virialized regions \citep{Lam13,Zu13}; 2) recovery of 3D phase space of galaxies from 2D measurements in the presence of projection effects \citep{Farahi16}; and \change{3) velocity bias which primarily impacts galaxies in clusters due to dynamical friction and baryonic effects \citep{Lau10,Munari13,Wu13,Anbajagane21}.}
In this paper, we provide an accurate model of the projected phase space structure of line-of-sight velocities. Further, we demonstrate that our model allows deriving unbiased estimates of critical halo properties for our clusters, including the halo edge radius, as well as the spatial and dynamical profiles of orbiting and infalling galaxies.

In detail, we use synthetic galaxy catalogs generated using the UniverseMachine \citep{Behroozi18} run on the Multi-Dark Planck 2 N-body simulation \citep{multidark} to motivate parametric models for the surface number density and line-of-sight velocity dispersions of three kinematically distinct populations of galaxies in and around halos: orbiting, infalling, and background.  As we demonstrate below, our model enables us to: 1) use the distinct dynamical signatures of orbiting, infalling, and background galaxies to statistically separate these three populations across all cluster radii using line-of-sight velocity data; 2) robustly infer the three-dimensional halo edge radius from line-of-sight velocity dispersion measurements; and 3) use the radial and velocity scales associated with the orbiting and infalling galaxy populations, to enable accurate detection of the edge radius. We note the possibility of using infalling galaxies for these purposes is particularly interesting, as infalling galaxies have spent little time in the halo environment, and are therefore more likely to be free from velocity bias due to the baryonic physics in the halo environment.

Our paper is laid out as follows.  We describe the simulation and mock galaxy catalog in \Cref{sec:Meth}. We explain the classification of orbiting and infalling galaxies and how they impact the velocity dispersion and cluster mass measurements in \Cref{sec:orb_inf}. We present our new model of the projected phase space structure of dark matter halos in \Cref{sec:Model}, and verify the validity of this model using the mock galaxy catalog in \Cref{sec:calibration}. We will then discuss applications of our model in \Cref{sec:discuss}. Our main findings are summarized in \Cref{sec:conc}.

\section{Methodology}\label{sec:Meth}

\subsection{Mock Catalogs}\label{sec:mock}

We analyze mock galaxy catalogs constructed using the {\it MDPL2} (Multi-Dark Planck) DM-only $N$-body simulation. The simulation was performed with the L-GADGET-2 code, a version of the publicly available cosmological code GADGET-2 \citep{Gadget}. The simulation has a box size of $1 {\rm \,Gpc/h}$, with a physical force resolution that decreases from $13{\rm \,kpc/h}$ at high-$z$ to $5{\rm \,kpc/h}$ at low-$z$.
The particle mass is $1.51\times10^9 M_{\odot}/h$ with $3840^3$ particles in the box. It assumes the {\it Planck} 2013 cosmology with $\Omega_m= 0.307$, $\Omega_{\Lambda}=0.693$, $\sigma_8=0.823$, and $H_0 = 68 {\rm \,km(s\,Mpc)}^{-1}$. The halos and subhalos are identified using the Rockstar 6D phase space halo finder \citep{Behroozi13Roc}, and the merger tree is built using the Consistent-Tree algorithm \citep{Behroozi13Con}. More details of the simulation can be found in \citet{multidark}.

The mock galaxy catalog is constructed using the UniverseMachine \citep{Behroozi18}, which pastes galaxies into halos and subhalos. \change{In this algorithm, the in-situ star formation rate is parameterized as a function of halo mass, halo assembly history, and redshift. Model galaxies are grafted directly onto halos and subhalos in the Rockstar merger trees from MDPL2 simulation. The stellar mass of the halo is then computed by integrating the star formation rates over the time steps of the simulation output, tracking the merger history of the halo. The algorithm forces the statistical properties of the resulting galaxy distribution to match the following observational data across cosmic times:} (i) stellar mass functions; (ii) cosmic star-formation rates and specific star-formation rates; (iii) quenched fractions; (iv) correlation functions for all, quenched, and star-forming galaxies; and (v) measurements of the environmental dependence of central galaxy quenching, using isolation criteria to identify centrals and a counts-in-cylinders-based quantification of $\sim5$~Mpc density.  These measurements are matched across a broad range of redshifts ($0<z<10$).

For this study, we restrict ourselves to dark matter halos with $M_{\rm 200m}>10^{14}M_{\odot}/h$, focusing on the high mass clusters.  We also restrict ourselves to galaxies with a stellar mass $M_*>10^9 M_{\odot}/h$. Orphan galaxies are necessary to correct for artificially disrupted subhalos in the simulations and are added by extrapolating the position and velocity of disrupted subhalos according to \citet{jiang_vdB14}.
Because we expect the systematic uncertainty in the model to increase near halo centers, where the fraction of orphan galaxies is high, we will ignore the cluster core when testing our model.

%%%%%%%%%%%%%%%%%%%%%%%%%%%%%%%%%%%%%%%%%%%%%%%%%

\subsection{Measurement}
\label{sec:richness}

Using the distant observer approximation, we select the $z$ axis of the simulation box as the line-of-sight.
Each halo has a central galaxy that shares the halo's position and velocity.
The projected radial distance between the cluster's central galaxy and any other galaxy is $R=\sqrt{(x-x_{\rm cen})^2+(y-y_{\rm cen})^2}$, and the 3D radial distance is $r=\sqrt{(x-x_{\rm cen})^2+(y-y_{\rm cen})^2+(z-z_{\rm cen})^2}$.  We will consistently use the variable $R$ for projected distances, and the variable $r$ for three-dimensional distances.
The relative line-of-sight velocity (LOS) of a simulated 
galaxy is
\begin{equation}
v_{\rm LOS} = (v_{z}-v_{z,{\rm cen}})+aH(z)d_{\rm com,LOS},
\end{equation}
where $v_{z}-v_{z,{\rm cen}}$ is the peculiar velocity of the galaxy with respect to the cluster, and the second term describes the velocity due to the expansion of the universe where $d_{\rm com, LOS}$ is the comoving distance between the cluster and the galaxy along the LOS. We apply a maximum velocity cut $|v_{\rm LOS}| < 3000\ {\rm km\,s}^{-1}$ for the galaxies used to study the velocity distribution. \change{\footnote{This velocity cut is equivalent to 3-5 times the velocity dispersion of the clusters involved in the studies, large enough to encompass all galaxies associated with clusters.}} Error estimates for all quantities we measure in the simulation are obtained using the jackknife method. Specifically, we split the simulation box into 100 different regions using the comoving $z$-coordinate. That is, halos with $z_{\rm cen}=[10(i-1),10i){\rm Mpc}/h$ in comoving units belongs to the $i$-th subsample. All the subsamples are then aggregated together to measure the mean profiles and jackknife errors.

\section{Classification of Orbiting and Infalling Galaxies}
\label{sec:orb_inf}

\begin{figure}
\centering
\includegraphics[width=0.35\textwidth]{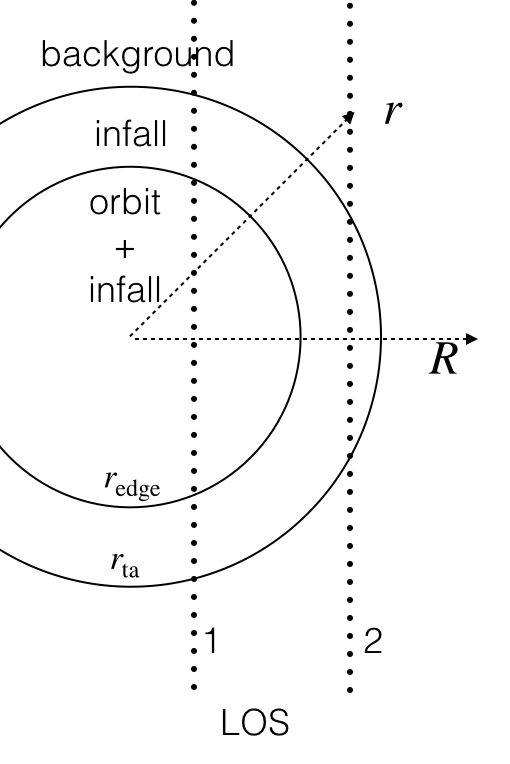}
\caption{A schematic diagram showing the split of the galaxy density field near a halo into its three components: orbiting galaxies, infalling galaxies, and background galaxies. The line of sight velocity dispersion profile for lines of sight near the halo center (e.g., LOS 1) receives contributions from all three types of galaxies, whereas more distant lines of sight (e.g., LOS 2) contain only infall and background galaxies.  The transition between these two limits occurs at the halo's edge radius.}
\label{fig:schematic}
\end{figure}

We classify the galaxies in the vicinity of a massive dark matter halo into 3 different categories: \textit{orbiting, infalling} and \textit{background} as follows. 
We define the turnaround radius $r_{\rm ta}$ to be the radius where the average physical radial velocity is zero ($\langle v_r\rangle$=0).  Galaxies at radial separations $r\geq \rta$ are defined to be background galaxies.
Galaxies with radial separations $r< \rta$ that have never experienced pericentric passages ($v_r<0$) inside the central halo are defined as infalling, while galaxies that have experienced pericentric passages are defined as orbiting. 
In 3D, the background galaxies are clearly separated from the orbiting and infalling components.  Between the halo edge radius --- defined as the maximal radial separation of the orbiting galaxies --- and the turnaround radius, one finds only infalling galaxies by definition.  Inside the edge radius, orbiting and infalling galaxies mix. A sketch of this setup is shown in \Cref{fig:schematic}.

\begin{figure}
    \centering
    \includegraphics[width=0.49\textwidth]{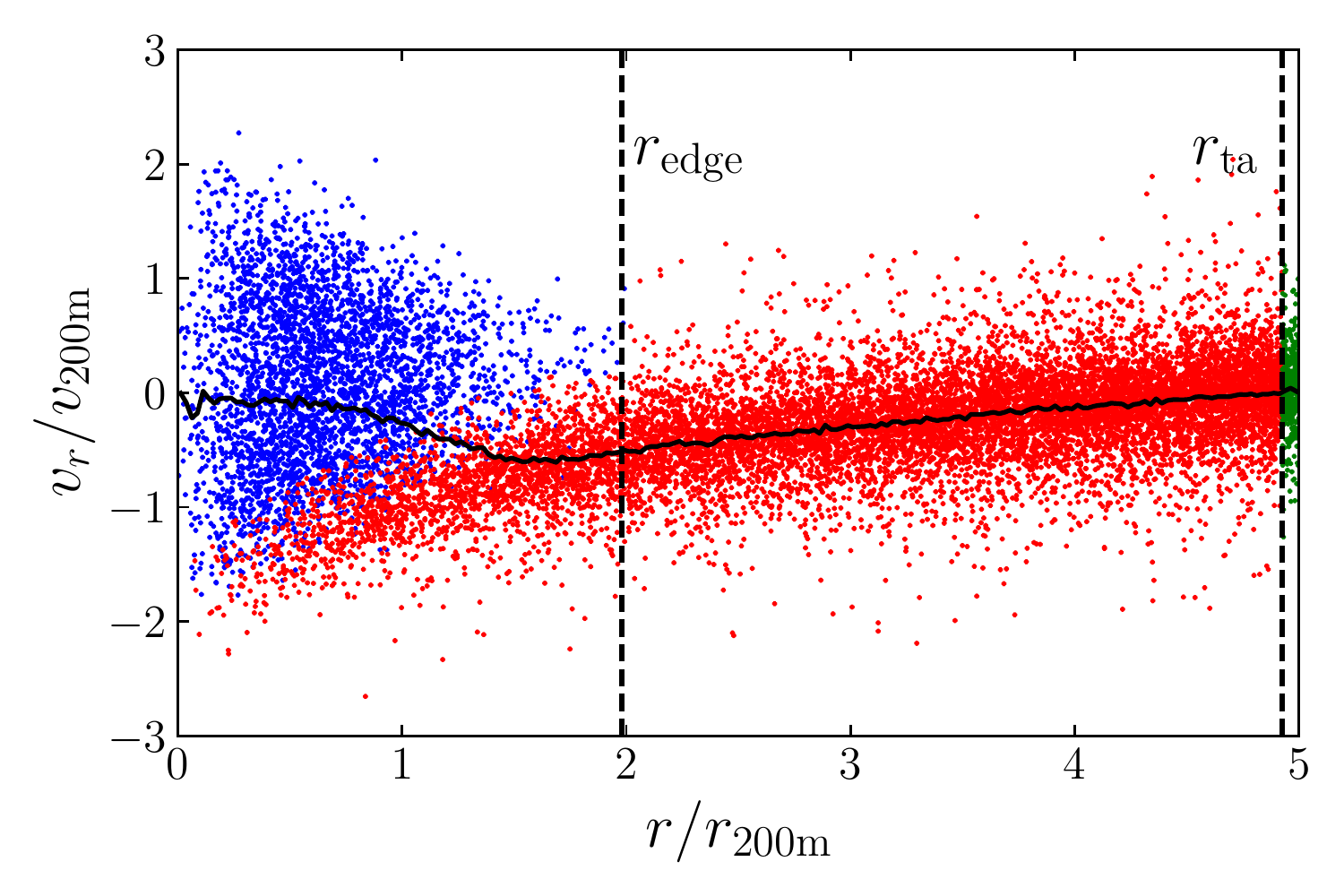}
    \caption{Peculiar radial velocity of galaxies as a function of radius. 
    Orbiting galaxies are denoted as blue,  infalling galaxies as red, and background galaxies as green.
    The halo edge radius is defined as the smallest radius containing all orbiting galaxies.
    Infalling galaxies start at the turnaround radius and continue all the way to the inner radii, with more negative radial velocities relative to the orbiting population. 
    Galaxies lie outside of the turnaround radius where the average radial velocity (black line) is larger than zero.
    }
    \label{fig:vr_r}
\end{figure}

Figure~\ref{fig:vr_r} shows the orbiting (blue) and infalling (red) galaxy populations in the $v_r$--$r$ plane, where $v_r$ is the physical radial velocity without the Hubble flow. Background galaxies are not shown because $\rta$ is larger than the radial range of the figure.  We see infalling galaxies have an ever-increasing inwards peculiar velocity while orbiting galaxies disperse around the zero radial velocity line.  The radial velocity dispersion of the orbiting galaxies decreases with increasing radius. We note that the infall stream protrudes deeply into the halo, well past the edge radius of the halo, and that the orbiting and infalling populations are mixed in phase space at small radii.

\begin{figure*}
    \centering
    \includegraphics[width=0.98\textwidth]{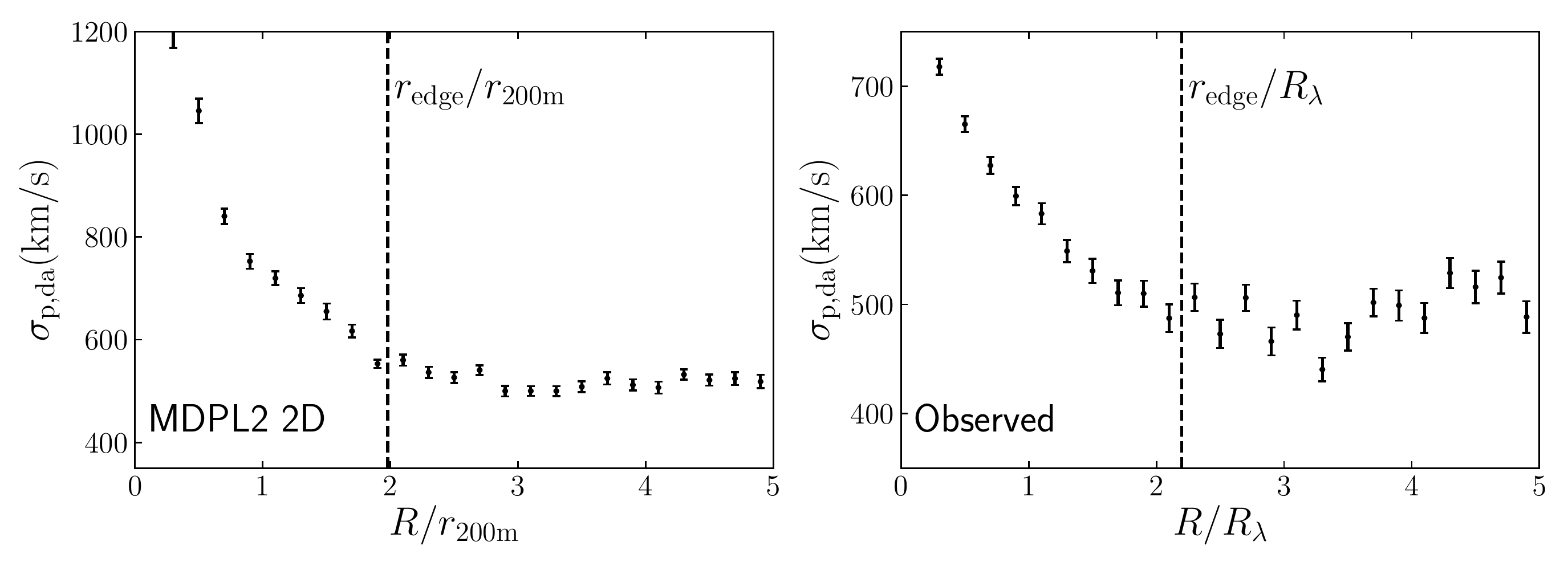}
    \caption{{\it Left panel:} Profile of the LOS velocity dispersion of dynamically associated galaxies (orbiting+infalling) as a function of projected radius in the mock catalog. The velocity dispersion monotonically decreases until $\Rout$, after which the profile flattens. {\it Right panel:} Fig 2 of \citet{Tomooka2020}, showing the "kink" in the observed dispersion profile similar to the feature shown in the  left panel, where $R_{\lambda}$ is the radius inside which velocity dispersion profile is increasing and outside which it is constant.}
    \label{fig:redge_2d}
\end{figure*}

We define the combination of orbiting and infalling populations around a halo as ``dynamically associated galaxies'' (indicated as ``da"), as their dynamics are significantly impacted by the gravity of the halo. \Cref{fig:redge_2d} shows the line-of-sight velocity dispersion profile of dynamically associated galaxies in simulation (left panel) and observation (right panel), as measured in \citet{Tomooka2020}. The two profiles share similar features: the dispersion profile first decreases with increasing radius, eventually flattening out past the presumed edge radius of the halo/cluster. One of the key goals of this work is to demonstrate that this feature can indeed be identified with the halo edge radius, as advocated in \citet{Tomooka2020}.

\section{A Model of Projected dark matter Phase Space Structures}
\label{sec:Model}

Given that we cannot classify which galaxies are orbiting, infalling or interlopers in the observed galaxy catalog, we improve upon the heuristic model of \citet{Tomooka2020} by characterizing the density and velocity dispersion profile of each type of galaxy in numerical simulations.   Specifically, we provide parameterizations that accurately describe the distribution of galaxy line-of-sight velocities as a function of projected radius, averaged over halos of fixed mass. We will leave halo-to-halo variations in these velocity profiles as a subject for future studies. Our model takes the form
\begin{align}
P(v,R)=& f_{\rm orb}(R) G(v,\sigma_{\rm orb}(R) ) + f_{\rm inf}(R) E(v,\sigma_{\rm inf}(R))]\nonumber\\ 
&+ (1-f_{\rm orb}(R)-f_{\rm inf}(R))G(v,\sigma_{\rm bg}),
 \label{eq:likelihood}
\end{align}
where $G(v,\sigma)$ --- used to describe both orbiting and background galaxies --- is a Gaussian distribution of mean zero and variance of $\sigma^2$.  \change{The infall population is modeled using the sech function, 
\begin{equation}
E(v,\sigma) = \frac{1}{2\sigma}\sech \left(\frac{\pi v}{2\sigma}\right),
\end{equation}
which has zero mean and variance of $\sigma^2$.} We will compare these distributions against the simulation data in \Cref{sec:calibration}.

To model the velocity distribution $P(v)$ using equation~\ref{eq:likelihood} we must also specify the width parameters $\sorb$, $\sinf$, and $\sbg$ which must depend on radius. We parameterize the radial profiles of the velocity dispersions as:
\begin{align}
\sigma_{\rm orb}(R) & =  \sigma_{\rm 0,orb} \left( q_{\rm orb} e^{-R/R_{\rm orb}}  + 1\right), \label{eq:sigmaorbprofile} \\
\sigma_{\rm inf}(R) & = \sigma_{\rm 0,inf} \left( q_{\rm inf} e^{-R/R_{\rm inf}} + 1 \right), \label{eq:sigmainfprofile}\\
\sbg(R) &= \sbg, \label{eq:sigmabgprofile}
\end{align}
where the amplitudes of the velocity dispersion profiles scale with mass as:
\begin{align}
\sigma_{\rm 0,inf} & = \sigma_{\rm p,inf}\left( \frac{M}{M_{\rm p}} \right)^{\alpha_{\rm inf}} ,\label{eq:sigmainf}\\
\sigma_{\rm 0,orb} & = \sigma_{\rm p,orb}\left( \frac{\sigma_{\rm 0,inf}}{\sigma_{\rm p,inf}} \right)^{\alpha_{\rm orb}} \label{eq:sigmaorb}\\
\sbg & = \sigma_{\rm p,bg}\left( \frac{\sigma_{\rm 0,inf}}{\sigma_{\rm p,inf}} \right)^{\alpha_{\rm bg}}, \label{eq:sigmabg}
\end{align} 
where $M_{\rm p}$ is a pivot mass, chosen to be $1.24\times10^{14}M_{\odot}/h$, the median mass of the halos we selected. $\sigma_{\rm p}$'s and $\alpha$'s are scaling parameters associated with the velocity dispersion of the orbiting and infalling populations. Note that in the above expressions we chose to scale the orbiting and background velocity dispersion with $\sigma_{\rm p,inf}$ rather than the halo mass. \change{$q_{\rm orb}$ and $q_{\rm inf}$ denote the ratio of velocity dispersion at the center of the cluster to the outskirt. We adopt these functions based on the features in the velocity dispersion profiles measured from the simulated data as explained below.}

This choice is motivated by the fact that the different scaling relations are driven by gravity and are therefore relatively robust to selection effects. If this is the case, then one can simply tie $\sigma_{\rm inf}$ to a cluster observable (mass, richness, SZ decrement, etc), while preserving the velocity dispersion scalings.  In other words, it is our hope that with this parameterization, the primary impact of cluster selection effects will be largely (though possibly not entirely) limited to the impact of the cluster scaling relation between $\sigma_{\rm inf}$ and the cluster observable. As to why we chose $\sigma_{\rm inf}$ as our base variable, as opposed to, say, $\sigma_{\rm orb}$, we anticipate that the relation between $\sigma_{\rm p,inf}$ and halo mass will be very robust to baryonic physics.  Of course, the relation between, say, the orbiting and infalling velocity dispersions could easily be impacted by baryonic physics, but this effect is now contained within a scaling relation that is directly observable.

The radial dependencies in the above equations are chosen based on the qualitative features of the profiles in simulations.  In particular, we found that $\sorb$ and $\sinf$ both decrease with increasing radius with a characteristic length scale $\Rorb$ and $\Rinf$.  Both profiles asymptote to a constant, and the relative amplitude of the constant ``shelf'' to the peak at small radii is characterized by the parameters $q_{\rm orb}$ and $q_{\rm inf}$. The amplitude of all three velocity dispersion profiles scale with mass either because (a) the galaxies are virialized, (b) the infall velocity is sensitive to the halo mass, or (c) the clustering bias at large scales is dependent on mass.

Finally, all the radial scales of the infalling and background galaxies scale with mass and/or velocity dispersion. We again use the infall velocity dispersion as our base variable, so that:
\begin{align}
\Rout &= \Routp  \left( \frac{\sigma_{\rm 0,inf}}{\sigma_{\rm p,inf}} \right)^{\alpha_{\rm edge}}, \label{eq:redge}\\
\Rorb &= \Rorbp  \left( \frac{\sigma_{\rm 0,inf}}{\sigma_{\rm p,inf}} \right)^{\alpha_{\rm rorb}},\label{eq:rorb}\\
\Rinf &= \Rinfp  \left( \frac{\sigma_{\rm 0,inf}}{\sigma_{\rm p,inf}} \right)^{\alpha_{\rm rinf}},\label{eq:rinf}
\end{align}
where $R_p$'s and $\alpha$'s are fitted parameters associated with the shape of velocity dispersion of the orbiting and infalling populations. The scaling relations in \Cref{eq:redge}--\Cref{eq:rinf} allow us to stack halos with different masses and provide a method to convert the velocity dispersions into physical scales.

The fraction of each type of galaxy is given by the ratio of the surface number densities ($\Sigma$): 
\begin{equation}
\label{eq:frac}
f_{\rm x} = \frac{\Sigma_x}{\surfaceorb +  \surfaceinf + \surfacebg},
\end{equation}
where $f_{\rm x}$ is the fraction of galaxies of type ``x", which can be orbiting (orb), infalling (inf), or background (bg). Note that the fraction $f_{\rm x}$ is invariant under a multiplicative constant being applied to all surface density profiles.  We use this invariance to arbitrarily normalize the orbiting profile so that $\rho_{\rm orb}=1$ at $r=0$.  Consequently,
the normalization parameters describing the infalling and background surface density profiles characterize only the normalization relative to $\Sigma_{\rm orb}$.

The {\it dimensionless} projected surface density of orbiting galaxies is given by:
\begin{equation}
\surfaceorb(R) = \int_{0}^{l_{\rm edge}} \rho_{\rm 3d}(r) \frac{{\rm d}l}{\Rout}, \label{eq:surorb}
\end{equation}
where $l = \sqrt{r^2-R^2}$ and $l_{\rm edge} = \sqrt{\max{\{\Rout^2-R^2,0}\}}$. The dimensionless 3D density profile of orbiting galaxies is modeled following \citet{diemer_kravtsov2014}:
\begin{equation}
\rho_{\rm 3d}(r)  = \exp\left(-\frac{2}{\alpha}\left[\left(\frac{r}{r_s}\right)^{\alpha}-1\right]  \right) 
\left[1+\left(\frac{r}{r_t}\right)^{\beta}\right]^{-\gamma/\beta}, \label{eq:3dorb}
\end{equation}
where \change{$\beta=4$ and $\gamma=6$ control the steepening of the slope near the splashback radius, and $\alpha=0.155+0.0095\nu^2$ determines how quickly the inner profile slope steepens. These parameter estimations are directly taken from the dark matter density profile fits from \citet{diemer_kravtsov2014}. The peak-height $\nu$ is defined as $\delta_c/\sigma(M,z)$, where $\delta_c$ is the critical overdensity collapse threshold, and $\sigma(M,z)$ is the variance of the linear density field on the scale of the halo. $r_s$ is the scale radius where the concentration of the halo $\r200m/r_s$ is given as a function of mass following \citet{duffy08}. } $r_t=\Rout/\rratio$, and $\rratio$ denotes the ratio of edge radius to the splashback radius as defined by steepest density slope.  We parameterize
$\surfaceinf$ and $\surfacebg$ as
\begin{align}
\surfaceinf &= s_1 (R/\Rout)^{s_2}, \label{eq:surinf}\\
\surfacebg &= s_3 (R/\Rout)^{s_4},\label{eq:surbg}
\end{align} 
where $s_1$ and $s_3$ provide the normalization of the infalling and background projected profiles relative to the orbiting contribution, and $s_2$ and $s_4$ provide the logarithmic slope respective to the projected radius. \change{The normalization for the projected profiles depends on the mass of the halo. We, thus, parametrize them as 
\begin{align}
s_1 & = s_{\rm 1p}\left( \frac{\sigma_{\rm 0,inf}}{\sigma_{\rm p,inf}} \right)^{\alpha_{\rm s1}},\label{eq:sinf}\\
s_3 & = s_{\rm 3p}\left( \frac{\sigma_{\rm 0,inf}}{\sigma_{\rm p,inf}} \right)^{\alpha_{\rm s3}}\label{eq:sbg}.
\end{align} 
We found that $s_2$ and $s_4$ do not have a strong mass dependence, and including them as additional parameters adds unnecessary degeneracy. Thus, we decide to ignore mass dependence.}

In addition, we define $\surfaceda$ as the surface number density of dynamically associated galaxies (infall+orbiting), and $\surfaceall$ as the surface number density of all (orbiting+infalling+background) galaxies. We emphasize we are not interested in highly accurate descriptions of the density profiles themselves.  Rather, we wish to achieve parametric descriptions that are ``good enough'' to accurately infer the fraction of orbiting/infalling galaxies as a function of radius, as this is the quantity that goes into our model.  We perform this comparison in \Cref{sec:calibration}.

\begin{figure*}
    \centering
    
    \tikzstyle{blockleft} = [rectangle, draw, fill=red!20,
    text width=10em, rounded corners, minimum height=4em]
    \tikzstyle{blockright} = [rectangle, draw, fill=red!20,
    text width=8em, rounded corners, minimum height=4em]
    \tikzstyle{blocktiny} = [rectangle, draw, fill=red!20,
    text width=5.5em, rounded corners, minimum height=4em]
    \tikzstyle{block} = [rectangle, draw, fill=red!20,
    text width=15em, rounded corners, minimum height=4em]
    \tikzstyle{line} = [draw, thick, -latex']
    
    \begin{tikzpicture} [node distance = 4cm, auto]
        \node [blocktiny] (obs) {
        \textbf{Mass Observables}\\
        $M$};
        \node [blockright, right of=obs, node distance = 2.5cm] (infall) {
        \textbf{Infall \\Dispersion}\\
        $\sigma_{\rm 0,inf}$ in \Cref{eq:sigmainf}};
        \node [blockleft, right of=infall, node distance = 3.5cm] (vscales) {
        \textbf{Velocity Scales}\\
        $\sigma_{\rm 0,orb},\sigma_{\rm bg}$ in  \Cref{eq:sigmaorb,eq:sigmabg}};
        \node [blockleft, below of=vscales, node distance = 2cm] (dscales) {
        \textbf{Distance Scales}\\
        $\Rout, \Rorb, \Rinf$ in  \Cref{eq:redge,eq:rorb,eq:rinf}};
        \node [block, right of=vscales, node distance = 4.5cm] (velocity) {
		\textbf{Velocity Dispersion Profiles}\\
        $\sigma_{\rm orb},\sigma_{\rm inf},\sigma_{\rm bg}$ in \Cref{eq:sigmaorbprofile,eq:sigmainfprofile,eq:sigmabgprofile}};
        \node [block, below of=velocity, node distance = 2cm] (fraction) {
        \textbf{Galaxy Distribution Profiles}\\
        $\Sigma_{\rm orb},\Sigma_{\rm inf},\Sigma_{\rm bg}$ in \Cref{eq:surorb,eq:3dorb,eq:surinf,eq:surbg}};
        
        \node [blockright, right of= velocity, node distance = 4.5cm] (dist) {
		\textbf{Distribution Function}\\
		$P(v,R)$ in \Cref{eq:likelihood}
        };
        \path [line] (obs) -- (infall);
        \path [line] (infall) -- (vscales);
        \path [line] (vscales) -- (velocity);
		\path [line] (infall) -- (dscales);
		\path [line] (dscales) -- (velocity);
        \path [line] (dscales) -- (fraction);
        \path [line] (fraction) -- (dist);
        \path [line] (velocity) -- (dist);
    \end{tikzpicture}
    \caption{
    The schematic diagram shows how the velocity distribution function is constructed along with the equations where these quantities are defined. The projected radii and velocities of the galaxies are normalized with cluster mass dependent distance and velocity scales. We then construct the normalized radial profiles of surface densities and velocity dispersions, which define the amplitude and width of the velocity distributions respectively. \Cref{tab:my_label} describes the parameters fitted for the model along with the associated physical quantities.}
    \label{fig:schematic_relations}
\end{figure*}
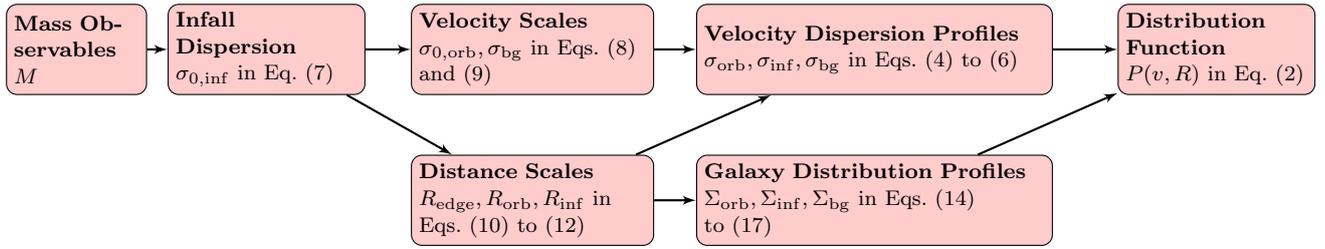

The summary of our model is described in \Cref{fig:schematic_relations}, and the free parameters associated with each of the quantities are listed in \Cref{tab:my_label}.

\begin{table}
   \centering
   \begin{tabular}{c|c|c}
       Key quantity & Parameters & Equations\\
       \hline
       $\Sigma$ & $k_t,s_{1p},s_2,s_{3p},s_4$ & \Cref{eq:surorb,eq:3dorb,eq:surinf,eq:surbg}\\
       $\sigma_{\rm inf}$ & $\alpha_{\rm inf},\sigma_{\rm p,inf},q_{\rm inf}$ & \Cref{eq:sigmainfprofile}\\
       $\sigma_{\rm orb}$ & $\alpha_{\rm orb},\sigma_{\rm p,orb},q_{\rm orb}$ & \Cref{eq:sigmaorbprofile}\\
       $\sigma_{\rm bg}$ & $\alpha_{\rm bg}$ & \Cref{eq:sigmabgprofile}\\
       $\Rout$ & $\Routp,\alpha_{\rm edge}$ & \Cref{eq:redge}\\
       $\Rorb$ & $\Rorbp,\alpha_{\rm rorb}$ & \Cref{eq:rorb}\\
       $\Rinf$ & $\Rinfp,\alpha_{\rm rinf}$ & \Cref{eq:rinf}
   \end{tabular}
   \caption{\change{The table lists the key physical quantities (surface densities $\Sigma$, velocity dispersions $\sigma$, and distance scales $R$), the free parameters associated with the quantities along with the equations that describe their relations. To see how the key quantities construct the final distribution function, see the summary in \Cref{fig:schematic_relations}.} For the results in \Cref{sec:calibration}, all parameters are fitted.}
    \label{tab:my_label}
\end{table}

\section{Model Validation}
\label{sec:calibration}

We now demonstrate that the model from \Cref{sec:Model} allows us to recover the 3-dimensional structure of orbiting, infalling, and background galaxies from the line-of-sight-projected observables in simulations.

The likelihood of the projected galaxy velocity data is 
\begin{equation}
    \lkhd = \prod_{i} P(v_i|R_i,M_i),
\end{equation}
where $P(v_i|R_i)$ is the probability distribution in equation~\ref{eq:likelihood}.  The product is
over all galaxies within a given projected radius of the cluster and within a velocity cut $v \leq 3,000\ {\rm km/s}$.  The variable $v_i$ is the line-of-sight velocity of galaxy $i$, $R_i$ is the projected radius of galaxy $i$, and $M_i$ is the mass of the central halo of galaxy $i$.  Note all these quantities are observables, with the exception of the halo mass. \change{For our analysis, we will use $M_{\rm 200m}$ (and $\r200m$) as known quantities from the halo catalog. But, when analyzing real data}, the halo mass would be replaced by an observable halo-mass proxy (e.g., richness, integrated SZ signal, or X-ray luminosity).  In other words, for the purposes of our analysis, we are treating halo mass only as a mass proxy: our analysis does not rely on the fact that our ``mass proxy'' is the bonafide halo mass.  We will characterize the impact of selection effects and the use of observable mass proxies on our results in future work.  For now, we restrict ourselves to demonstrating that in the absence of such complications, our modeling framework accurately describes the velocity profile of halos.%}.

\begin{figure}
    \centering
    \includegraphics[width=0.49\textwidth]{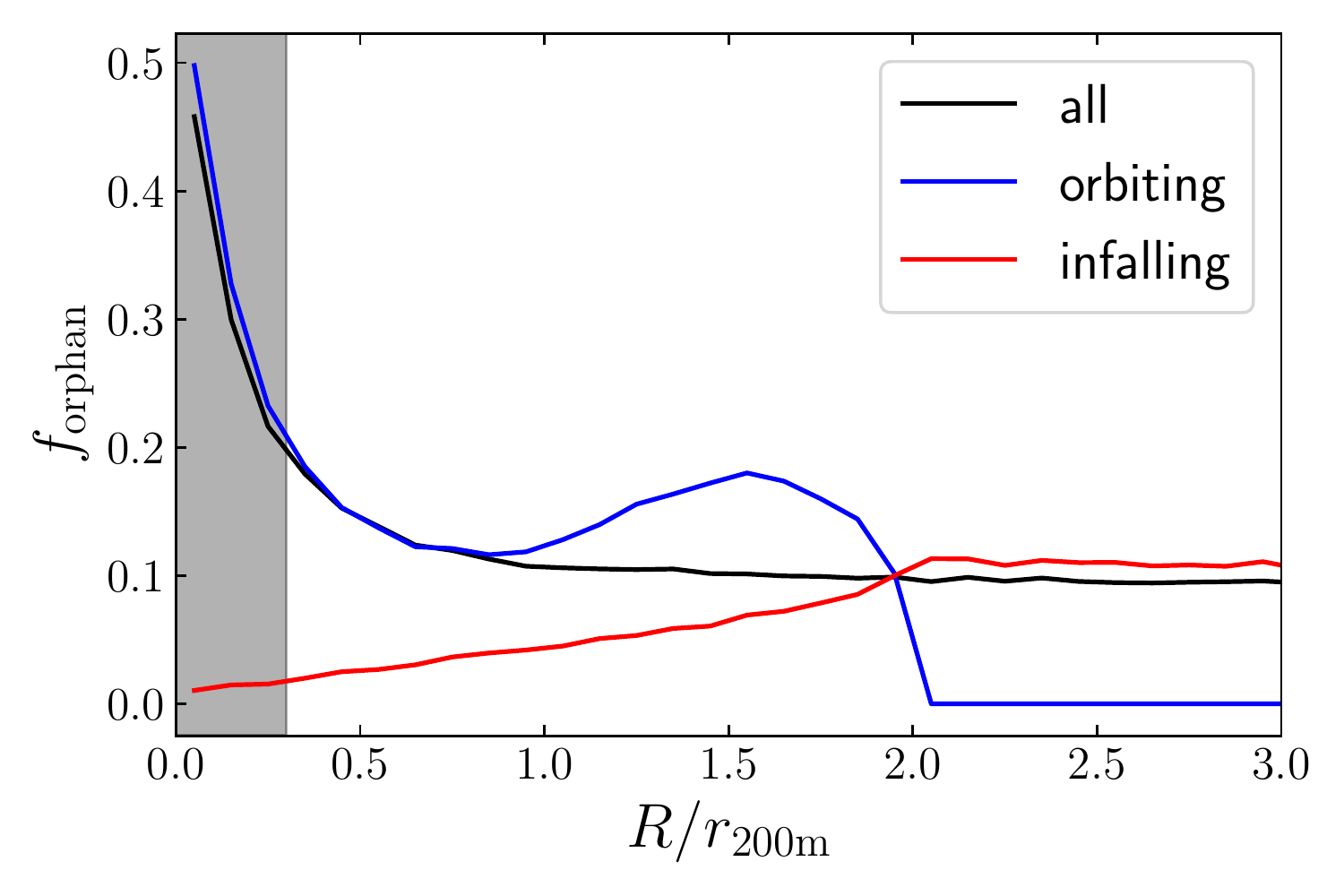}
    \caption{Fraction of the number of orphan galaxies to the total number of galaxies within each sample, all (black), orbiting (blue), or infalling (red) galaxies, as a function of the 2D projected radius. Most of the orphan galaxies are orbiting galaxies residing within $r<0.3r_{\rm 200m}$ (indicated with the grey shaded region). The orphan fraction of orbiting galaxies rises around $1.0<r/r_{\rm 200m}<1.5$, as the total number of orbiting galaxies decrease drastically. The infalling population, on the other hand, increases monotonically with radius and asymptotes to a constant value of $0.12$ at $r/r_{\rm 200m}>2.0$.}
    \label{fig:orphan}
\end{figure}

When analyzing data, we ignore all galaxies in the cluster's central region ($R\lesssim 0.3r_{200m}$). The fraction of orphan galaxies in the simulation increases towards the halo center, rising from $\approx 15\%$ at $R=0.3r_{\rm 200m}$ to $40\%$ at $R<0.05 r_{\rm 200m}$ as shown in \Cref{fig:orphan}. By restricting our analysis to radii $R\geq 0.3r_{\rm 200}$, \change{we drastically reduce the impact of orphan galaxies on the galaxy velocity dispersion and distribution profiles which vary by no more than 3\% with or without orphans.} We note that baryonic physics can also change the velocity dispersion of the galaxies within the 3D radius $r\lesssim0.2\r200m$ \citep{Lau10}, so our $R\geq 0.3r_{\rm 200m}$ cut should also help in this regard.

We use the MCMC sampler Emcee \citep{emcee} to create a realization of the posterior of our model parameters given a simulated dataset. We adopt the maximum likelihood point as our best-fit model. During the fit, we applied a prior such that all radii and the surface number densities were positive. Furthermore, we demand $\sigma_{\rm p,bg}>\sigma_{\rm p,orb(0)}+500\ {\rm km/s}$ to distinguish between the Gaussian of the background from that of the orbiting galaxies in the fit. We checked that our results are robust when changing $500$ to $[300-1000]{\rm km/s}$.

\begin{figure*}
\centering
\includegraphics[width=0.98\textwidth]{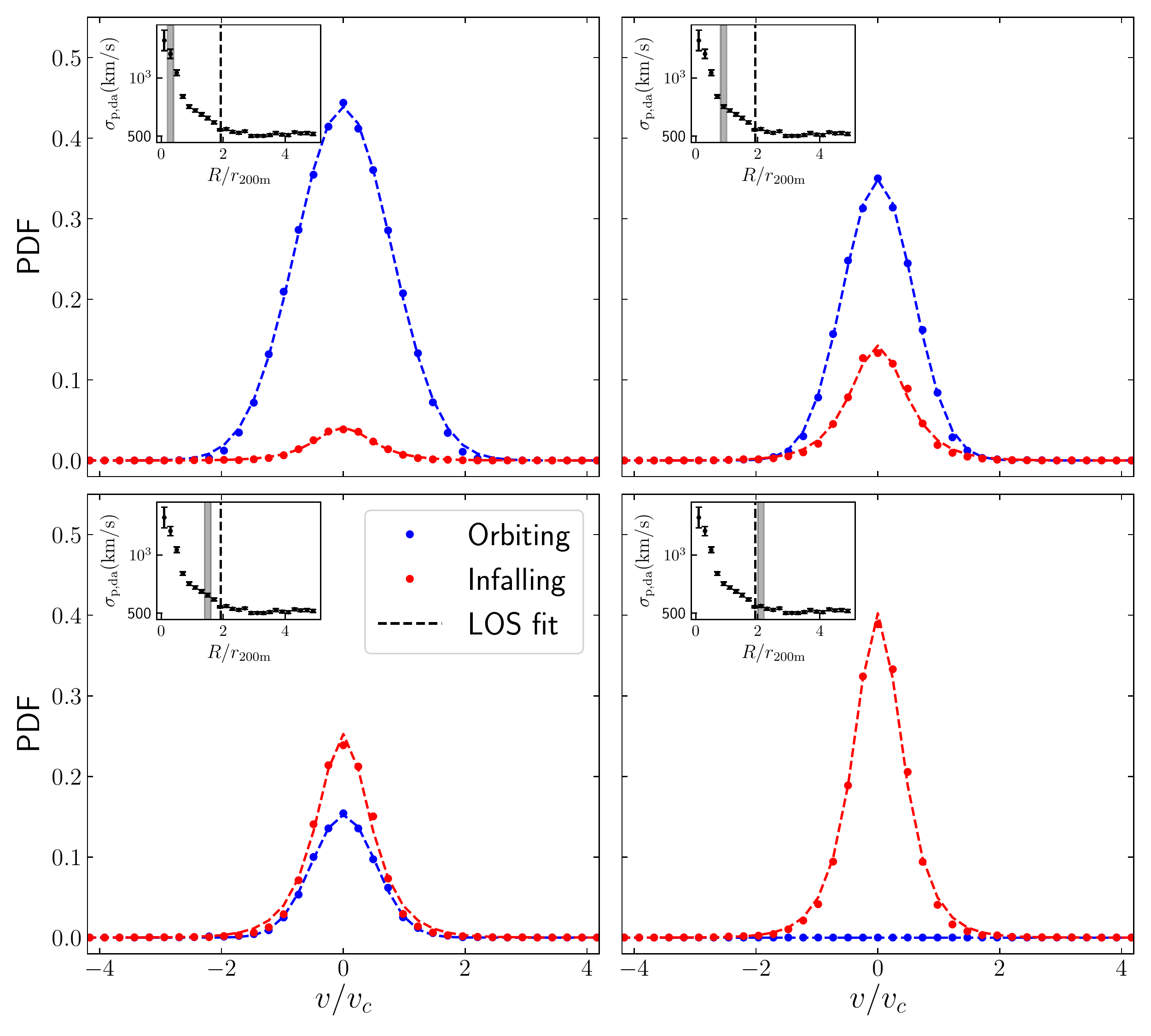}
\caption{The velocity distributions of the orbiting and infalling components using all galaxies along the line-of-sight (LOS) for four 2D radial shells of $0.2<R/\r200m<0.4$ (top left panel), $1.1<R/\r200m<1.3$ (top right panel), $1.6<R/\r200m<1.8$ (bottom left panel) and $2.0<R/\r200m<2.2$ (bottom right panel). Note that the first 3 panels correspond to LOS 1 and the bottom right panel to LOS 2 in the schematic diagram in \Cref{fig:schematic}. The horizontal axis is scaled by $v_c=\sqrt{GM_{\rm 200m}/r_{\rm 200m}}$. The counts are normalized for the total population in each radial bin separately. The lines indicate the distributions obtained by using the fitting procedure described in \Cref{sec:Model}, \change{which relies exclusively on projected data, while the dots are the distributions of orbiting and infalling galaxies based on the pericentric passage, which themselves require the full dynamical history of every galaxy in the sample.
The insets show the velocity dispersion profile of orbiting and infalling galaxies combined, along with a shaded band indicating the radial bin of the galaxies shown in the panel.} Our model allows us to statistically recover the distribution of orbiting and infalling galaxies from the projected dynamical data alone. 
}
\label{fig:distribution}
\end{figure*}

\Cref{fig:distribution} compares the velocity distribution $P(v|R)$ obtained for orbiting and infalling galaxies using our likelihood model (dashed curves) to that inferred from splitting galaxies into orbiting, infalling, and background galaxies using 3D information and particle orbits as described in \Cref{sec:orb_inf}. The figure is restricted to 4 representative radial bins, though we emphasize we fit for the velocity distribution at all radii $R<3r_{\rm 200m}$ simultaneously. The model parameters that we varied for this chain are listed in \Cref{tab:my_label}, and include all the parameters introduced in our model. We see that our parametric model provides an excellent fit to the data across all cluster radii.  We emphasize that the velocity distributions inferred from our model rely exclusively on observables, and do not benefit from an a priori split of galaxies into orbiting and infalling.  Note too the lack of an orbiting galaxy population at radii larger than the halo edge radius. 

\begin{figure}
    \centering
    \includegraphics[width=0.49\textwidth]{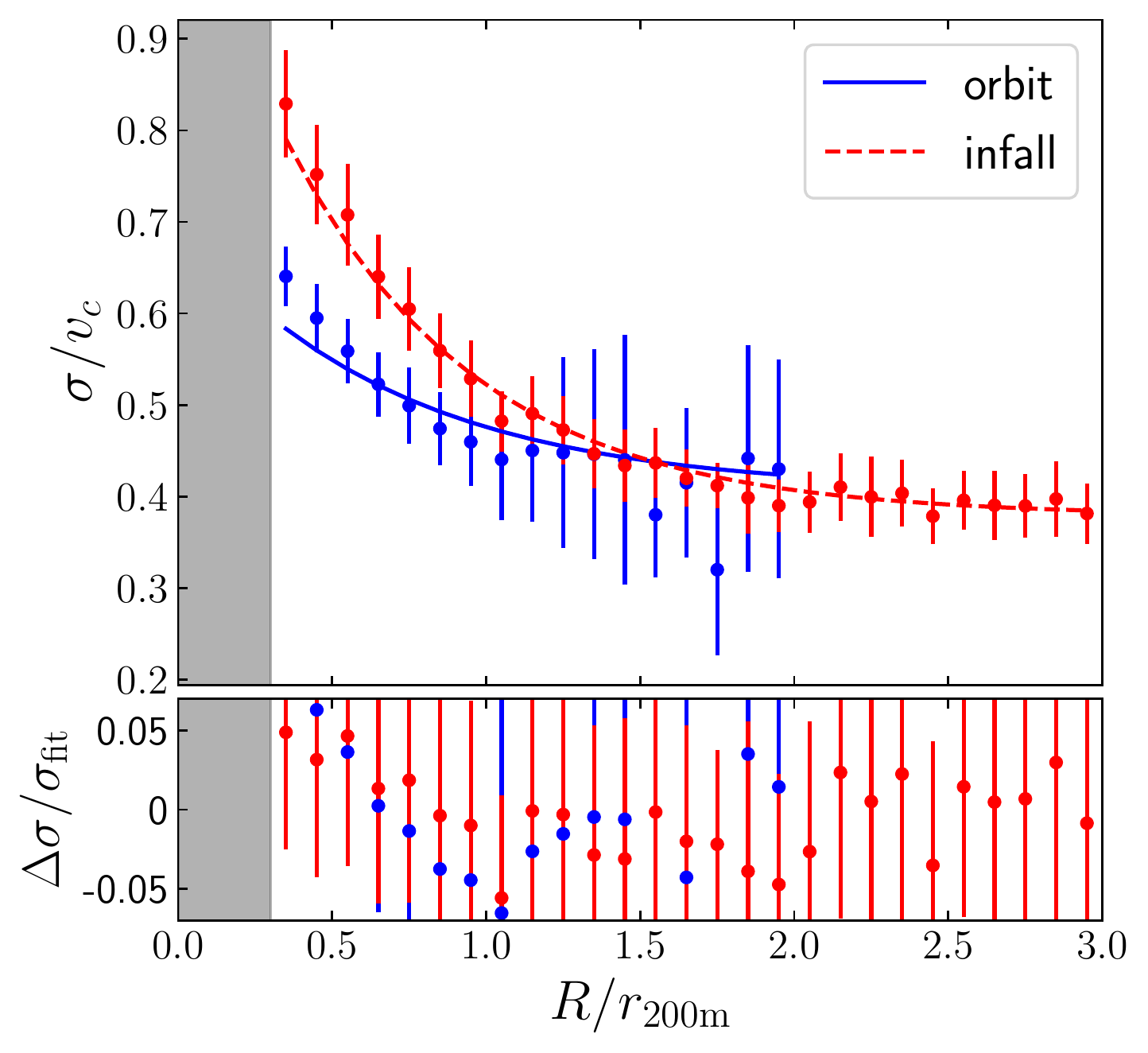}
    \caption{
    The stacked LOS dispersion profile of orbiting (blue) and infalling (red) galaxies as a function of 2D projected radius for dark matter halos. The data points are measured from the stacked profile, while the errorbars are obtained with jackknife resampling. The bottom panel shows that the best-fit model recovers the velocity dispersion profiles of the orbiting populations. The shaded region has a large fraction of orphan galaxies, and is not included in our fits.
    }
    \label{fig:dispersion_vir}
\end{figure}

The lines in \Cref{fig:dispersion_vir} show our best fit model for the LOS velocity dispersion of orbiting and infalling galaxies as a function of the projected radius. The data points are the dispersions estimated using orbiting and infalling galaxies only, as tagged based on their orbital properties. The number of degrees of freedom is hard to estimate, since the fit is done at the likelihood level using individual galaxies.  We set the degrees of freedom to the number of points (17 and 27 for orbiting and infalling, respectively) minus the number of parameters describing the velocity profile (5 parameters each for the orbiting ($\sigma_{\rm orb},R_{\rm orb}$) and the infalling ($\sigma_{\rm inf},R_{\rm inf}$) populations in \Cref{tab:my_label}). The corresponding $\chi^2/dof$ are $10.4/12$ and $17.3/22$ for orbiting and infalling galaxies, respectively. That is, our parametric model provides a statistically acceptable description of the data.

\begin{figure}
    \centering
    \includegraphics[width=0.49\textwidth]{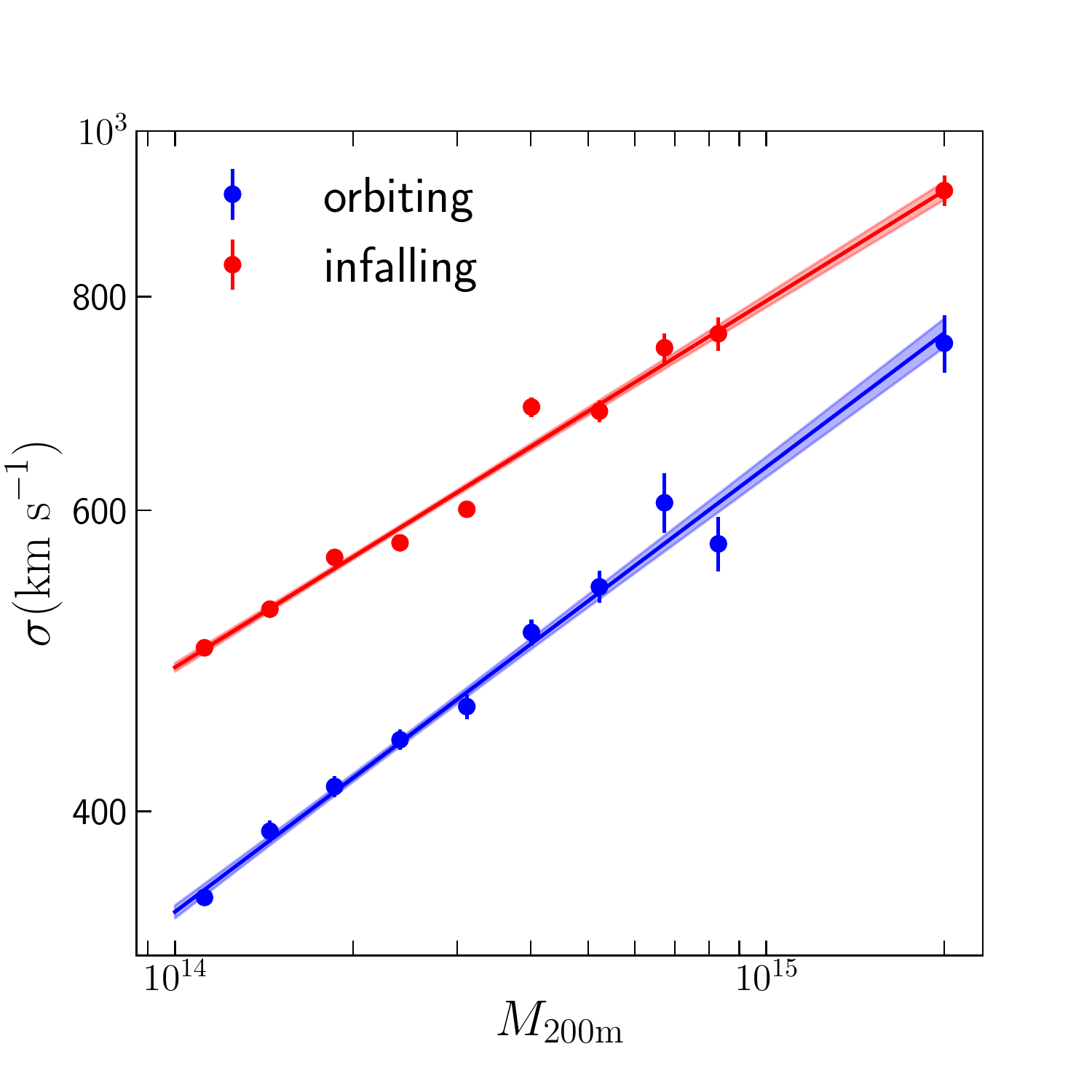}
    \caption{The line-of-sight velocity dispersion-mass relation of the orbiting and infalling galaxies. The data points in the figure represent the values fitted from the stacked profile in logarithmically spaced mass bins.  The errorbars represent the error on the fitted values. The line indicates the median best-fit relation. The shaded region indicates $(16-84)$-percentile around the best-fit relation. }
    \label{fig:dispersion_mass}
\end{figure}

We further demonstrate that our analysis recovers the correct scaling between the mass and each of the three radial scales, $\Rout$, $\Rinf$, and $\Rorb$ and velocity dispersion scales $\sigma_{\rm orb}$ and $\sigma_{\rm inf}$. To do so, we repeat our analysis using only halo--galaxy pairs for halos in narrow mass bins.  For this analysis, we modify our model so that $\Rout$, $\Rinf$, $\Rorb$, $\sigma_{\rm orb}$ and $\sigma_{\rm inf}$ are constants within each individual mass bin.  \Cref{fig:dispersion_mass} shows the recovered values of $\sigma_{\rm inf}$ and $\sigma_{\rm orb}$ in each mass bin, along with the best-fit relation from our global model. The errors are determined from the posterior distribution of the fit in each mass bin. The resulting data points are well fit using the proposed power-law model.

\begin{figure}
    \centering
    \includegraphics[width=0.48\textwidth]{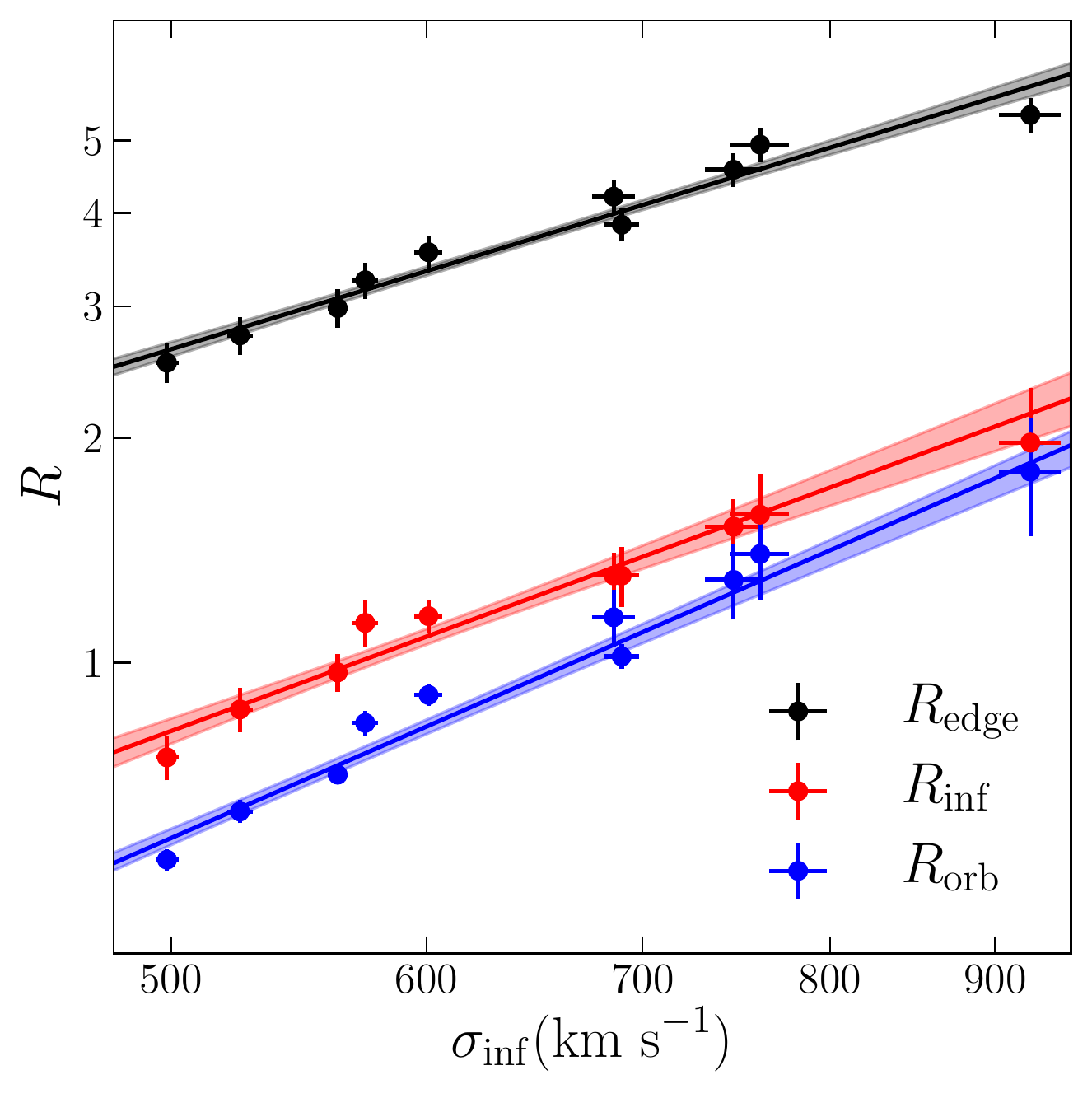}
    \caption{The scaling relations of the radial scales ($\Rout, \Rorb, \Rinf$) in units of Mpc as a function of the LOS velocity dispersion of infalling galaxies. The data points in the figure represent the values fitted from the stacked profile in logarithmically spaced mass bins.  The errorbars represent the error on the fitted values. The line indicates the median best-fit relation. The shaded region indicates $(16-84)$-percentile around the best-fit relation.
    }
    \label{fig:r_sigma}
\end{figure}

\Cref{fig:r_sigma} shows the recovered values of $\Rout$, $\Rinf$, $\Rorb$ instead. We see again that the resulting data points are well fit using the proposed power-law model. The calibrated amplitude and slopes of the three relations given the pivot point $1.24\times10^{14}M_{\odot}/h$ are
\begin{align}
    \log_{10} \Routp =  0.505\pm0.006,\ & \,\alpha_{\rm edge} = 1.32\pm 0.07,\\
    \log_{10} \Rorbp = -0.113\pm0.009,\ & \,\alpha_{\rm rorb} = 1.89\pm 0.05,\\
    \log_{10} \Rinfp = 0.0111\pm0.006,\ & \,\alpha_{\rm rinf} = 1.60\pm 0.08,
\end{align}
where the radii are in units of Mpc.  

\begin{figure}
    \centering
    \includegraphics[width=0.48\textwidth]{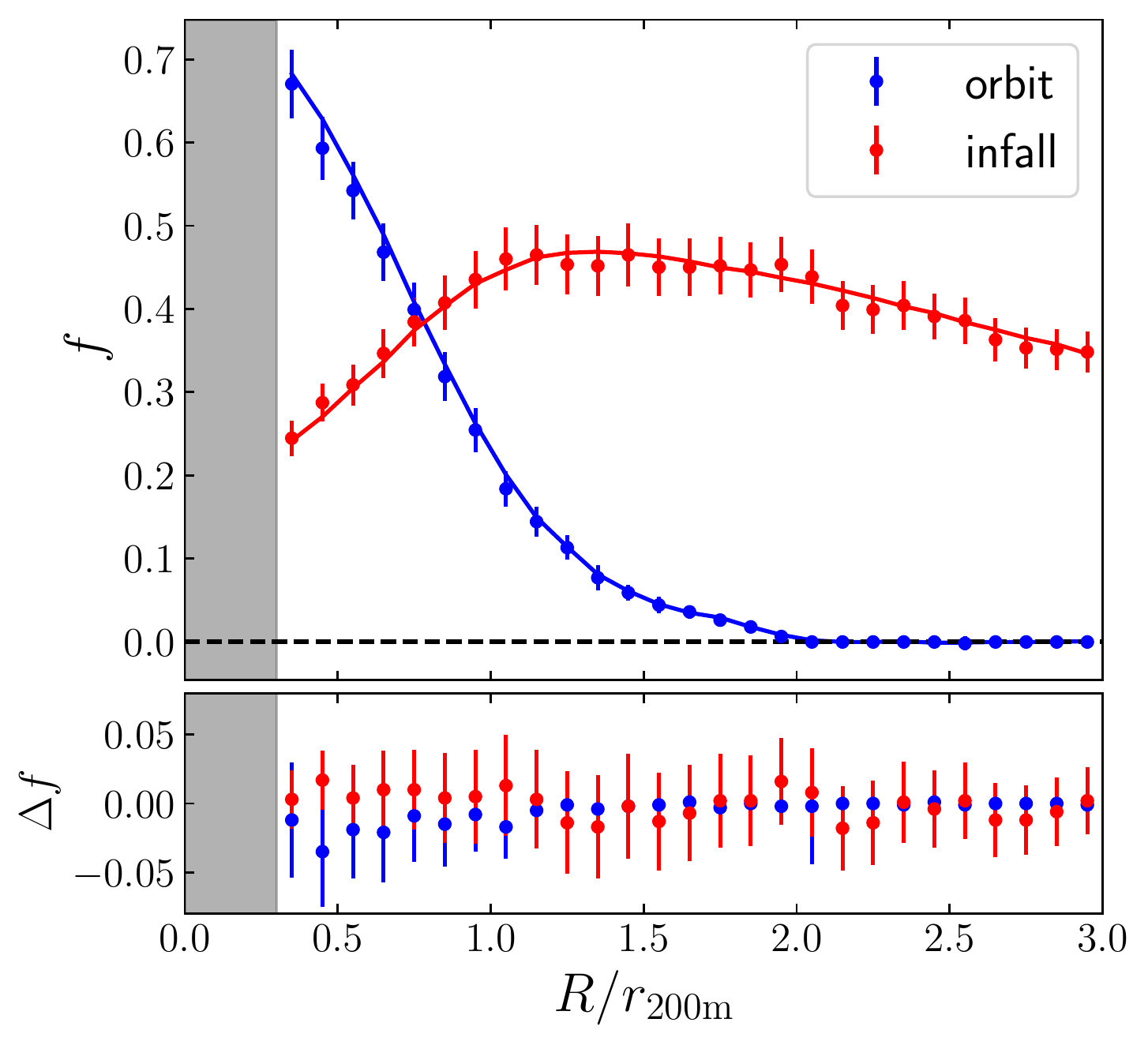}    
    \caption{ The fraction of the orbiting, and infalling galaxies as a function of projected radius. The data points indicate the measurements from the simulations and the errorbars are obtained with jackknife resampling. The lines indicate the best-fit model for each population. Our fits do not include the shaded region due to the large fraction of orphan galaxies in the cores of halos. These fractions along with the interloper fraction add up to 1, where the interloper fraction increases as a function of the projected radius.
    }
    \label{fig:fraction}
\end{figure}

Finally, we turn to examine how the fraction of orbiting and infalling galaxies changes as a function of the halo-centric projected radius $R$. \Cref{fig:fraction} compares the fractions inferred from our model to those obtained using the orbiting/infalling split described in \Cref{sec:orb_inf}.  As expected, the fraction of orbiting galaxies decreases monotonically with radius, and approaches 0 at $\Rout$. The infall fraction, on the other hand, increases with radius up to  $R\lesssim 1.2 r_{\rm 200m}$ (due to the decreasing number of orbiting galaxies), but then decreases as we move further from the halo as expected. Once again, our best fit model provides an excellent match to the data, with a $\chi^2/dof$ of $11.23/14$ and $16.48/23$ for the orbiting and infalling profiles.  As before, the number of degrees of freedom is set to the number of data points (17 for orbiting and 27 for infall and background) minus the number of parameters describing each of surface densities ($\Sigma, R_{\rm edge}$ in \Cref{tab:my_label}).

\section{Implications for Hubble Constant Measurements with Galaxy Clusters}\label{sec:discuss}

In this work, we identify three dynamically different populations of galaxies, along with three length scales that can be used to infer cluster distances: $\Rout$, $\Rorb$, and $\Rinf$. Our measurements rely on the fact that each of these three radial scales is tightly correlated with the observed velocity dispersions. In our model, we have chosen to use $\sinf$ as our ``base'' variable upon which other variables depend, but we could have also picked $\sorb$. The model described above has the potential to self-calibrate systematic uncertainties in the proposed analysis. For example, baryonic effects and dynamical friction are expected to be small for the dynamics of infalling galaxies which have not yet experienced environmental effects in the virialized region of clusters. We, therefore, expect $\sinf$ to be much less susceptible to the non-linear cluster astrophysics than $\sorb$.  Thus, the comparison of the edge radius and the other radii scales obtained using multiple observables, such as $\sinf$ and $\sorb$, may enable us to self-calibrate systematic uncertainties associated with baryonic physics in our proposed measurement. We briefly describe below a possible application of our modeling framework, which we intend to pursue in future work.

\begin{figure}
    \centering
    \includegraphics[width=0.48\textwidth]{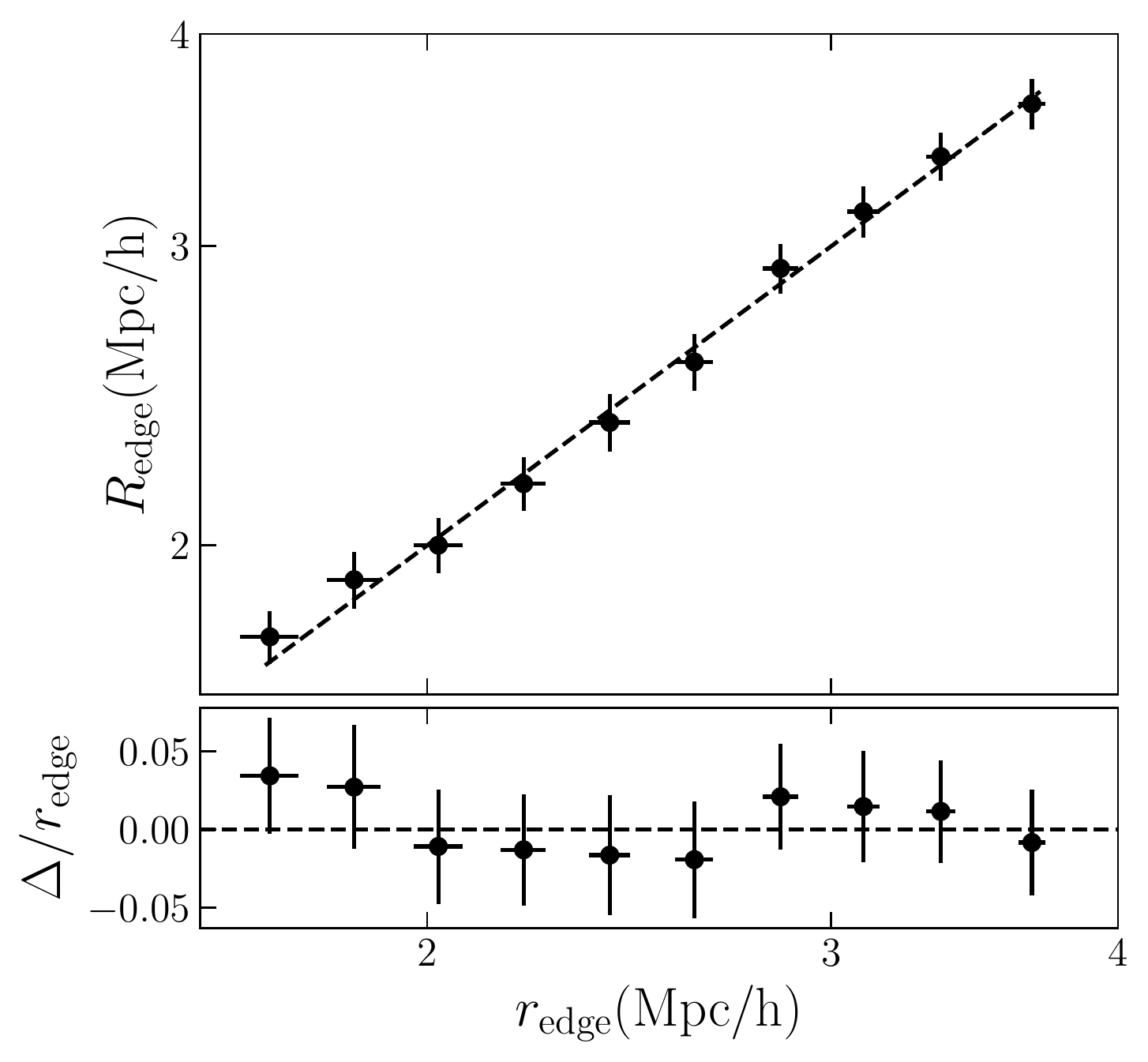}
    \caption{The relation between the three-dimensional edge radius inferred from particle orbits ($\rout$) and the projected edge radius as inferred from the line-of-sight velocity data ($\Rout$). The data points in the figure represent the values from logarithmically spaced mass bins in the range $10^{14}<M_{\rm 200m}/(M_{\odot}/h)<10^{15}$. The errorbars on 3D edge radius represent the jackknife error, while the errorbars in the inferred radii are obtained from the posterior. The dashed line indicates the 1-to-1 relation.}
    \label{fig:redge_redge}
\end{figure}

\citet{Wagoner2020} argue that: 1) the feature in the velocity dispersion profile of galaxy clusters corresponds to the halo edge radius of the clusters; and 2) that this feature can be used as a standard ruler.  Specifically, the predicted error in dimensionless Hubble constant $h = H_0/100{\rm km\,s}^{-1}$ is
\begin{equation}
\sigma_h = \sqrt{\left(\frac{h_{\rm fid}\sigma_{R_{\rm p,edge}}}{R_{\rm p,edge}}\right)^2 + \sigma_{\rm h,fid}^2},\label{eq:huncertainty}
\end{equation}
where $\sigma_{R_{\rm p,edge}}$ is the error on the simulation calibrated pivot edge radius and $\sigma_{\rm h,fid}$ is the statistical uncertainty in the recovered cluster edge radius.   The latter depends on the specific survey assumptions adopted in the forecast, and was estimated by \citet{Wagoner2020} to be $\approx 0.009$ in DESI. To determine the validity of the proposal by \citet{Wagoner2020}, we first check whether the edge radius recovered in the projected phase space from our model $\Rout$ coincides with the 3D edge radius $\rout$. The 2D edge radius comes from our likelihood analysis using narrow mass bins.  The error in $\Rout$ is obtained from the posterior of the fit.  The 3D edge radius is the smallest radius containing all orbiting galaxies.  Its error is obtained by jackknifing the simulation box. \Cref{fig:redge_redge} compares the recovered edge radius $\Rout$ to the true, three-dimensional radius $\rout$ determined using the orbiting/infall/background split described in \Cref{sec:orb_inf}. Evidently, our model allows us to correctly recover the three-dimensional edge radius $\rout$, validating the \citet{Wagoner2020} proposal.  Given the calibrated value of $\sigma_{R_{\rm p,edge}}/R_{\rm p,edge} = 0.0138$ and $h_{\rm fid} = 0.67$, the calibration floor on the Hubble parameter when using our simulations is $\sigma_h=0.0129$.

In addition to edge radius, our method provides three distinct distance scales that can be used to constrain the Hubble parameter. Thus, if we also fix these additional distance scales, we can obtain a tighter constraint.
To test this, we generate a mock galaxy catalog where the projected radii are converted to angular distances assuming $h_{\rm fid} = 0.67$. We then fix the scaling relations of all three radial scales $\Rout,\Rinf,$ and $\Rorb$ and fit for the Hubble parameter. We find that the error on the Hubble parameter is reduced by a factor of 2 when using all 3 distance scales, relative to the constraints derived using only $R_{\rm edge}$.

Naively, the calibration floor in our analysis is larger than the statistical uncertainty of DESI. However, \citet{Wagoner2020} assume that the scaling relations between each of our three radii $\Rout$, $\Rorb$, and $\Rinf$, and the velocity dispersions $\sorb$ and $\sinf$, are cosmology independent. However, if the relations between $R$--$\sigma$ depend on the Hubble parameter $h$, the previously estimated error can either increase or decrease. From the virial theorem, we naively expect $\sigma^2 \propto M/R$.  Traditional spherical-overdensity halo definitions take the form of $M\propto \Delta \rho_{\rm c}R^3$ where $\rho_{\rm c}$ is the critical density of the Universe and $\Delta$ is the average overdensity enclosed within the sphere with a radius $R$.  Since $\rho_{\rm c}\propto h^2$, we can combine the two equations to arrive at $R\propto h^{-1}\sigma$. If this were the case, then the combination $R(\sigma)/D_A$ would in fact be independent of $h$, effectively ruining the entire argument. However, it is important to emphasize that: 1) our base velocity dispersion is that of \it infalling \rm galaxies, so the use of the virial theorem is questionable, and 2) the above prediction relies on traditional spherical overdensity halo definitions.  These definitions are unphysical, and fail to capture the dynamics of orbiting structures in a halo.  Indeed, notice that the above argument also implies that the slope of the $R$--$\sigma$ relation is one and that of $R$--$M$ is one-third.  By contrast, the mass-radius relations of the splashback and other similar radius definitions defining the outer boundary do not follow $M\propto R^3$ \citep{more_etal2015,garcia2020}, as the overdensity of the splashback $\Delta$ introduces additional mass dependence \citep{diemer_etal17,shi_2016a}. 
Equations~\ref{eq:redge} through \ref{eq:rinf} have slopes in the range $1.3$ to $1.9$, hinting that the dependence on $h$ may not be that of our naive prediction. For our analysis, we will assume that $R_x\propto h^{-1\pm \alpha_h}\sigma_{\rm 0,inf}^{\alpha_x}$, where $\alpha_h$ indicates deviation from the virial theorem. Then \Cref{eq:huncertainty} can be rewritten as 
\begin{equation}
\sigma_h = \sqrt{\left(\frac{h_{\rm fid}\sigma_{R_{\rm p,edge}}}{R_{\rm p,edge}}\right)^2 + \left(\frac{\sigma_{\rm h,fid}}{\alpha_h }\right)^2}.\label{eq:huncertainty_hdependence}
\end{equation}
Given the same calibrated value of $\sigma_{R_{\rm p,edge}}/R_{\rm p,edge} = 0.0138$, $h_{\rm fid} = 0.67$, and $\sigma_{\rm h,fid} = 0.009$, for $\alpha_h=1$, we recover the original uncertainty with $\sigma_h=0.0129$. To determine $\alpha_h$, we employ the Quijote simulations \citep{Quijote} with $h=0.6511,0.6711,0.6911$. We then select halos with mass above $10^{14}M_{\odot}/h$ as clusters, and randomly subsample the dark matter particles by a factor of 10 to generate a mock sample of ``galaxies.''  We find $R_{\rm edge}$ in each of the three simulations, and fit our results to a power-law in $h$.  We find that $\alpha_h=0.78\pm 0.10$, with a $2.5\%$ error on the Hubble parameter. Taking into account the fact that including additional distance scales will reduce the error by a factor of 2, we expect an error of $\approx1.3\%$ on the Hubble parameter with the data from the DESI survey.

\section{Conclusions}\label{sec:conc}

We present a new parametric model of the projected velocity field of massive dark matter halos, and test it on a mock galaxy catalog produced with UniverseMachine run on the MDPL2 N-body simulation.  Our model splits galaxies into three kinematically distinct populations: orbiting, infalling, and background.  Each population has a different radially-dependent velocity distribution and surface density profile.   Our main findings are as follows:

\begin{enumerate}    
    \item Our model accurately describes the radially dependent line-of-sight velocity distribution of dark matter halos.  In particular, using only projected observables, a likelihood analysis based on our model correctly recovers the velocity distribution of each of the three types of galaxies: orbiting, infalling, and background (see \Cref{fig:distribution}).  
  
    \item In addition to using $\Rout$, the smallest radius containing all orbiting galaxies, the halo velocity dispersion profile contains two additional length scales $\Rorb$ and $\Rinf$.  These two radii govern the drop-off of the orbiting and infalling velocity dispersions as a function of radius (see \Cref{fig:dispersion_vir}), and represent two additional clusters scales that can be empirically recovered from the projected velocity dispersion of galaxy clusters.
    
    \item The amplitudes of the velocity dispersion profiles $\sorb$ and $\sinf$ are correlated with the cluster scales $\Rorb$, $\Rinf$, and $\Rout$ (see \Cref{fig:r_sigma}). Calibration of these scaling relations enables us to use these scales as standard rulers, thereby enabling us to measure the Hubble parameter $h$ from projected cluster observables.
    
    \item Our model allows us to recover unbiased estimates of the fraction and dispersion of orbiting/infalling galaxies as a function of radius (see \Cref{fig:fraction}).  This is, to our knowledge, the first method capable of statistically distinguishing between these two galaxy populations using observable data. 
    
\end{enumerate}

Our work provides the first step toward modeling the projected phase space structure beyond the virial and splashback radius, while taking into account projection effects.  We caution that many additional sources of systematic uncertainty remain to be characterized (e.g., miscentering, selection functions, and baryonic effects).
 
We intend to address each of these in turn in future works, with the goal of realizing a percent-level measurement of the projected velocity distribution using upcoming spectroscopic surveys, such as DESI, PFS, and SPHEREx.

\section*{Acknowledgement}
\change{We thank the anonymous referee for helpful comments on the draft.}
The CosmoSim database used in this paper is a service by the Leibniz-Institute for Astrophysics Potsdam (AIP). The MultiDark database was developed in cooperation with the Spanish MultiDark Consolider Project CSD2009-00064. 
HA and DN acknowledge support from Yale University and the facilities and staff of the Yale Center for Research Computing.  DN and ER also acknowledge funding from the Cottrell Scholar program of the Research Corporation for Science Advancement. ER and BW are supported by DOE grants DE-SC0015975 and DE-SC0009913. ER is also supported by NSF grant 2009401. 

\section*{Data Availability}
The MDPL2 halo catalogs and the mock UM galaxy catalogs are publicly available at https://www.peterbehroozi.com/data.html.

\bibliography{ref}
\bibliographystyle{mnras}

\end{document}